\documentclass[twoside,twocolumn,journal]{article}
\usepackage[latin9]{inputenc}
\usepackage{verbatim}
\usepackage{amsmath}
\usepackage{amssymb}
\usepackage{graphicx}

\makeatletter
\@ifundefined{date}{}{\date{}}

\usepackage{ltexpprt}

\newtheorem{definition}{Definition}[section]

\newtheorem{problem}{Problem}[section]

\makeatother

\begin{document}

\title{Traffic Control for Network Protection Against Spreading Processes}

\author{Victor M. Preciado, Michael Zargham, and David Sun%
\thanks{The authors are with the Department of Electrical and Systems Engineering
at the University of Pennsylvania, Philadelphia PA 19104.%
}}
\maketitle
\begin{abstract}
Epidemic outbreaks in human populations are facilitated by the underlying
transportation network. We consider strategies for containing a viral
spreading process by optimally allocating a limited budget to three
types of protection resources: (\emph{i}) \emph{Traffic control resources},
(\emph{ii}), \emph{preventative resources} and (\emph{iii}) \emph{corrective
resources.} Traffic control resources are employed to impose restrictions
on the traffic flowing across \emph{directed} edges in the transportation
network. Preventative resources are allocated to nodes to reduce the
probability of infection at that node (e.g. vaccines), and corrective
resources are allocated to nodes to increase the recovery rate at
that node (e.g. antidotes). We assume these resources have monetary
costs associated with them, from which we formalize an optimal budget
allocation problem which maximizes containment of the infection. We
present a polynomial time solution to the optimal budget allocation
problem using \emph{Geometric Programming} (GP) for an arbitrary \emph{weighted
and directed }contact network and a large class of resource cost functions.
We illustrate our approach by designing optimal traffic control strategies
to contain an epidemic outbreak that propagates through a real-world
air transportation network.
\end{abstract}

\section{Introduction}

Designing strategies to contain spreading processes in networks is
a central problem in epidemiology and public health \cite{Bai75},
computer viruses \cite{GGT03}, as well as security of cyberphysical
systems \cite{roy2012security}. We consider the problem of containing
an epidemic outbreak in a \emph{weighted, directed} transportation
network through the allocation of a fixed budget. The budget can be
invested into three types of resources: (\emph{i}) \emph{Traffic control
resources}, (\emph{ii}) \emph{preventative resources,} and (\emph{iii})
\emph{corrective resources.} Traffic control resources are employed
to impose restrictions on the traffic flowing across (possibly directed)
edges in the transportation network. Preventative resources (i.e.
vaccines) are allocated to nodes to reduce the probability of infection
at that node, and corrective resources (i.e. antidotes) are allocated
to nodes to increase the recovery rate at that node. These resources
have monetary costs associated with them, from which we formalize
an optimal budget allocation problem which maximizes containment of
the infection for a fixed budget. Although our primary focus is on
epidemic control, the proposed framework is also relevant in distribution
of resources to control many other spreading processes, such as the
propagation of malware in computer networks or rumors in online social
networks.

The structure of spatial interaction (social contact topology) of
the population plays a key role in epidemics (e.g., SARS \cite{watts2005multiscale}).
There are several models of spreading mechanisms for arbitrary contact
networks in the literature. A common feature in the analysis of spreading
models is the \emph{basic reproduction ratio}%
\footnote{The basic reproductive ratio is defined as the average number of secondary
infections produced during an infected individual's infectious period,
when the individual is introduced into a population where everyone
is healthy \cite{AM91}. %
}. The analysis of this question in arbitrary contact networks was
first studied by Wang et al. \cite{WCWF03} for a Susceptible-Infected-Susceptible
(SIS) discrete-time model. In \cite{GMT05}, Ganesh et al. studied
the \emph{epidemic threshold}%
\footnote{A threshold on the infection rate to curing rate ratio, above which
epidemic occurs.%
} in a continuous-time SIS spreading processes. Similar analyses have
also been performed for a wide variety of spreading models \cite{CWWLF08,MOK09}.
Most papers in this area conclude that the spectral radius of the
network (i.e., the largest eigenvalue of its adjacency matrix) plays
a key role in the behavior of the spreading process.

The main problem studied in this paper can be stated as follows: Given
a contact network (possibly weighted and/or directed), resources which
impose restrictions on the the contact network, and resources that
provide partial protection (e.g., vaccines and/or antidotes), how
should one allocate a fixed budget to these resources to achieve maximal
containment? Considering only the protection resources, this problem
has been addressed through a variety of heuristics. Cohen et al. proposed
a heuristic vaccination strategy called \emph{acquaintance immunization
policy} which is much more efficient than random vaccine allocation,
\cite{cohen2003efficient}. \textbf{ }Borgs et al. studied theoretical
limits in the control of spreads in undirected network with a non-homogeneous
distribution of antidotes, \cite{BCGS10}. Chung et at. studied a
heuristic immunization strategy based on the PageRank vector of the
contact graph, \cite{chung2009distributing}. Wan et al. present a
control theory based strategy for undirected networks using eigenvalue
sensitivity analysis, \cite{WRS08}. This work is related to \cite{GOM11},
where the level of infection is minimized given an undirected network
and a fixed budget with which to allocate corrective resources. In
\cite{PZEJP13}, a convex formulation to find the optimal allocation
of protective resources in an undirected network using geometric programming
(GP). Linear-fractional programs are used in a similar context to
compute optimal disease awareness strategies to contain spreading
processes in social networks, \cite{PDS13}.

In this paper, we propose a convex framework to find the cost-optimal
distribution of traffic, preventive and corrective control resources.
To our knowledge, this work is the first to simultaneous allocation
of heterogeneous resources on nodes and edges. Furthermore, our approach
produces an exact solution to the allocation problem--without relaxations
or heuristics-- in polynomial time. Furthermore, this work applies
to the case of \emph{weighted and directed} networks with \emph{nonidentical}
agents.

We organize our exposition as follows. In Section \ref{Notation},
we introduce notation and background needed in our derivations. We
introduce a stochastic model to simulate viral spreading in Subsection
\ref{sub:Epidemic-Model}, and state the resource allocation problems
considered in this paper in Subsection \ref{sub:Problem-Statements}.
In Section \ref{sec:Convex Framework}, we propose a convex optimization
framework to efficiently solve the allocation problems in polynomial
time. Subsection \ref{sub:GP for SC Digraphs}, present the solution
to the allocation problem for strongly connected graphs. We illustrate
our results using a real-world air transportation network in Section
\ref{sub:Numerical-Results}. We include some conclusions in Section
\ref{sec:Conclusions}.

\section{\label{Notation}Preliminaries \& Problem Definition}

In the rest of the paper, we denote by $\mathbb{R}_{++}^{n}$ the
set of $n$-dimensional vectors with positive entries. We denote vectors
using boldface and matrices using capital letters. $I$ denotes the
identity matrix and $\mathbf{1}$ the vector of all ones. $\Re\left(z\right)$
denotes the real part of $z\in\mathbb{C}$.

\subsection{\label{sub:Graph-Theory}Graph-Theoretical Background.}

A \emph{weighted}, \emph{directed} graph (also called digraph) is
defined as $\mathcal{G}:=\left(\mathcal{V},\mathcal{E},\mathcal{W}\right)$,
where $\mathcal{V}:=\left\{ v_{1},\dots,v_{n}\right\} $ a set of
nodes, $\mathcal{E}\subseteq\mathcal{V}\times\mathcal{V}$ is a set
of directed edges, and $\mathcal{W}$ is an edge weight function $\mathcal{W}:\mathcal{E}\rightarrow\mathbb{R}_{++}$.
We define the in-neighborhood of node $v_{i}$ as $\mathcal{N}_{i}^{in}:=\left\{ j:\left(v_{j},v_{i}\right)\in\mathcal{E}\right\} $
and the weighted \emph{in-degree} of node $v_{i}$ as $\deg_{in}\left(v_{i}\right):=\sum_{j\in\mathcal{N}_{i}^{in}}\mathcal{W}\left(\left(v_{j},v_{i}\right)\right)$.
A directed path of length $l$ from $v_{i_{0}}$ to $v_{i_{l}}$ is
an ordered set of vertices $\left(v_{i_{0}},v_{i_{1}},\ldots,v_{i_{l}}\right)$
such that $\left(v_{i_{s}},v_{i_{s+1}}\right)\in\mathcal{E}$ for
$s=0,\ldots,l-1$. A directed graph $\mathcal{G}$ is \emph{strongly
connected} if, for every pair of nodes $v_{i},v_{j}\in\mathcal{V}$,
there is a directed path from $v_{i}$ to $v_{j}$.

The \emph{adjacency matrix} of a weighted, directed graph $\mathcal{G}$,
denoted by $A_{\mathcal{G}}=[w_{ij}]$, is a $n\times n$ matrix defined
entry-wise as $A_{ij}=\mathcal{W}((v_{j},v_{i}))$ if $(v_{j},v_{i})\in\mathcal{E}$,
and $A_{ij}=0$ otherwise. We only consider graphs with positively
weighted edges; hence, the adjacency matrix of a graph is always nonnegative.
Conversely, given a $n\times n$ nonnegative matrix $A$, we can associate
a directed graph $\mathcal{G}_{A}$ such that $A$ is the adjacency
matrix of $\mathcal{G}_{A}$. Finally, a nonnegative matrix $A$ is
\emph{irreducible} if and only if its associated graph $\mathcal{G}_{A}$
is strongly connected.

Given a $n\times n$ matrix $M$, we denote by $\mathbf{v}_{1}\left(M\right),\ldots,\mathbf{v}_{n}\left(M\right)$
and $\lambda_{1}\left(M\right),\ldots,\lambda_{n}\left(M\right)$
the set of eigenvectors and corresponding eigenvalues of $M$, respectively,
where we order the eigenvalues in decreasing order of their real parts,
i.e., $\Re\left(\lambda_{1}\right)\geq\Re\left(\lambda_{2}\right)\geq\ldots\geq\Re\left(\lambda_{n}\right)$.
We call $\lambda_{1}\left(M\right)$ and $\mathbf{v}_{1}\left(M\right)$
the dominant eigenvalue and eigenvector of $M$, respectively. The
spectral radius of $M$, denoted by $\rho\left(M\right)$, is the
maximum modulus of an eigenvalue of $M$.

\textbf{}%

\subsection{\label{sub:Epidemic-Model}Stochastic Modeling of Epidemic Outbreaks.}

Introduced by Weiss and Dishon in \cite{weiss1971asymptotic}, the
susceptible-infected-susceptible (SIS) model is a popular stochastic
model to simulate spreading processes. Wang et al. \cite{WCWF03}
proposed a discrete-time extension of the SIS model to simulate spreading
processes in networked populations. Van Mieghem et al. proposed in
\cite{MOK09} a continuous-time version, called the N-intertwined
SIS model, and rigorously analyze the connection between the speed
of spreading and the spectral radius of the contact network. In our
work, we consider a recent extension to Van Mieghem's model proposed
in \cite{VO13}. This model, called the heterogeneous N-intertwined
SIS model (HeNiSIS), presents the flexibility of allowing a heterogenous
distribution of agents in a networked population.

This HeNiSIS model is a continuous-time networked Markov process with
nodes being in one out of two possible states, namely, susceptible
(S) or infected (I). The state of node $v_{i}$ evolves according
to a stochastic process parameterized by a node-dependent infection
rate $\beta_{i}$, a node-dependent recovery rate $\delta_{i}$, and
the rate of contact between $v_{i}$ and its neighbors. These contact
rates are quantified by the set of weights $\left\{ w_{ij},j\in\mathcal{N}_{i}^{in}\right\} $.
The main novelty of the HeNiSIS model is that it allows to consider
the effect of a heterogeneous distribution of parameters throughout
the contact network. In Section \ref{sub:Problem-Statements}, we
shall assume that $\beta_{i}$, $\delta_{i}$, and $w_{ij}$ are adjustable
by allocating protection resources to the nodes and edges of the directed
contact graph.

In the HeNiSIS model, the state of node $v_{i}$ at time $t\geq0$
is a binary random variable $X_{i}\left(t\right)\in\{0,1\}$. The
state $X_{i}\left(t\right)=0$ (resp., $X_{i}\left(t\right)=1$) indicates
that node $v_{i}$ is in the susceptible (resp., infected) state.
We define the vector of states as $X\left(t\right):=\left(X_{1}\left(t\right),\ldots,X_{n}\left(t\right)\right)^{T}$.
{} Using Kolmogorov forward equations and a mean-field approach, one
can approximate the dynamics of the spreading process using a system
of $n$ ordinary differential equations, as follows. Let us define
$p_{i}\left(t\right):=\Pr\left(X_{i}\left(t\right)=1\right)=E\left(X_{i}\left(t\right)\right)$,
i.e., the marginal probability of node $v_{i}$ being infected at
time $t$. Hence, the Markov differential equation \cite{van2006performance}
for the state $X_{i}\left(t\right)=1$ is the following,
\begin{equation}
\frac{dp_{i}\left(t\right)}{dt}=\left(1-p_{i}\left(t\right)\right)\beta_{i}\sum_{j=1}^{n}w_{ij}p_{j}\left(t\right)-\delta_{i}p_{i}\left(t\right).\label{eq:HeNiSIS dynamics}
\end{equation}
Considering $i=1,\ldots,n$, we obtain a system of nonlinear differential
equation with a complex dynamics. In the following, we derive a sufficient
condition for infections to die out exponentially fast. This ODE presents
an equilibrium point at $\boldsymbol{p}^{*}=0$, called the disease-free
equilibrium. A stability analysis of this ODE around the equilibrium
provides the following stability result \cite{PZEJP13}:

\begin{proposition} \label{prop:Heterogeneous SIS stability condition}Consider
the nonlinear HeNiSIS model in (\ref{eq:HeNiSIS dynamics}) and assume
that $A_{\mathcal{G}}\geq0$ (entry-wise nonnegative), and $\beta_{i},\delta_{i}>0$.
Then, if 
\begin{equation}
\Re\left[\lambda_{1}\left(\mbox{diag}\left(\beta_{i}\right)A_{\mathcal{G}}-\mbox{diag}\left(\delta_{i}\right)\right)\right]\leq-\varepsilon,\label{eq:Spectral Control}
\end{equation}
for some $\varepsilon>0$, then $\left\Vert \boldsymbol{p}\left(t\right)\right\Vert \leq\left\Vert \boldsymbol{p}\left(0\right)\right\Vert K\exp\left(-\varepsilon t\right)$,
for some $K>0$.

\end{proposition} 

In the proof of Proposition \ref{prop:Heterogeneous SIS stability condition}
in \cite{PZEJP13}, we showed that the linear dynamical system $\dot{\boldsymbol{p}}\left(t\right)=\left(BA_{\mathcal{G}}-D\right)\boldsymbol{p}\left(t\right)$
upper-bounds the mean-field approximation in (\ref{eq:HeNiSIS dynamics});
thus, the spectral result in (\ref{eq:Spectral Control}) is a sufficient
condition for an initial infection to die out exponentially fast in
the HeNiSIS model. Therefore, we can use the above proposition to
find an allocation of resources able to shift the real parts of the
eigenvalues of $\mbox{diag}\left(\beta_{i}\right)A_{\mathcal{G}}-\mbox{diag}\left(\delta_{i}\right)$
to the complex right half-plane.

\subsection{\label{sub:Problem-Statements}Traffic Control Problem.}

Our objective in this paper is to control the spreading of a viral
outbreak by distributing protection resources throughout the nodes
and edges of a contact network. We consider three types of protection
resources:
\begin{itemize}
\item Traffic-control resources resources that can be allocated to the (directed)
edges of the network. This resource can be used to imposes restrictions
on the traffic flowing across edges of the transportation network.
In particular, allocating this resource to edge $\left(v_{j},v_{i}\right)\in\mathcal{E}$
has the effect of reducing the weight of that edge in a prescribed
interval $w_{ij}\in\left[\underline{w}_{ij},\overline{w}_{ij}\right]$,
where $\underline{w}_{ij}$ and $\overline{w}_{ij}$ are the minimum
and maximum feasible weights for edge $\left(v_{j},v_{i}\right)$.
\item Preventive resources, which can be allocated to the nodes in the network
to modify the infection rates. This resource, when allocated to node
$v_{i}$, have the effect of reducing the infection rate in the feasible
range $\beta_{i}\in\left[\underline{\beta}_{i},\bar{\beta}_{i}\right]$.
\item Corrective resources which can be used to increase the recovery rate
of node $v_{i}\in\mathcal{V}$ in the range $\delta_{i}\in\left[\underline{\delta}_{i},\bar{\delta}_{i}\right]$.
\end{itemize}
We consider these protection resources have an associated cost. We
define three cost functions: (\emph{i}) Traffic-control cost functions
$h_{ij}\left(w_{ij}\right)$ for $\left(v_{j},v_{i}\right)\in\mathcal{E}$,
(\emph{ii}) vaccination cost functions $f_{i}\left(\beta_{i}\right)$,
and (\emph{iii}) antidote cost functions $g_{i}\left(\delta_{i}\right)$,
for $i\in\mathcal{V}$. In the rest of the paper, we assume that the
traffic control function and the vaccination cost function, $h_{ij}\left(w_{ij}\right)$
and $f_{i}\left(\beta_{i}\right)$, are monotonically decreasing w.r.t.
$\beta_{i}$ and $w_{ij}$. Also, we assume the antidote cost function
$g_{i}\left(\delta_{i}\right)$ to be monotonically increasing w.r.t.
$\delta_{i}$.

In this paper we solve the \emph{budget-constrained} allocation problem:
Given a total budget $C$, find the best allocation of vaccines, antidotes
and traffic-control resources to maximize the exponential decay rate
of $\left\Vert \boldsymbol{p}\left(t\right)\right\Vert $. Based on
Proposition \ref{prop:Heterogeneous SIS stability condition}, the
decay rate of an epidemic outbreak is determined by $\varepsilon$
in (\ref{eq:Spectral Control}); hence, we maximize $\varepsilon$
(the decay rate) such that $\left\Vert \boldsymbol{p}\left(0\right)\right\Vert K\exp\left(-\overline{\varepsilon}t\right)$.

In mathematical terms, we formulate the budget-constrained allocation
problem as follows:

\begin{problem}\label{Problem: Budget Constrained Allocation}\emph{Given
the following elements:}
\begin{enumerate}
\item \emph{A (positively) weighted, directed network $\mathcal{G}=\left(\mathcal{V},\mathcal{E},\mathcal{W}\right)$,}
\item \emph{A set of cost functions $\left\{ f_{i}\left(\beta_{i}\right),g_{i}\left(\delta_{i}\right)\right\} _{v_{i}\in\mathcal{V}}$
and $\left\{ h_{ij}\left(w_{ij}\right)\right\} _{(v_{j},v_{i})\in\mathcal{E}}$,}
\item \emph{Bounds on the infection, recovery, and traffic rates $0<\underline{\beta}_{i}\leq\beta_{i}\leq\overline{\beta}_{i}$,
$0<\underline{\delta}_{i}\leq\delta_{i}\leq\overline{\delta}_{i}$,
$i\in\mathcal{V}$,} \emph{and $0<\underline{w}_{ij}\leq w_{ij}\leq\overline{w}_{ij}$,
}$(v_{j},v_{i})\in\mathcal{E}$,\emph{ and}
\item \emph{A total budget $C$,}
\end{enumerate}
\emph{find the cost-optimal distribution of traffic control resources,
vaccines and antidotes to maximize the exponential decay rate $\varepsilon$.}

\end{problem}

In what follows we state the problem in terms of traffic flows $w_{ij}$,
infection and recovery rates, $\beta_{i}$, $\delta_{i}$. Let us
define the matrix decision variables $W_{\mathcal{G}}:=\left[w_{ij}\right]$,
$B=\mbox{diag}\left(\beta_{i}\right)$, and $D:=\mbox{diag}\left(\delta_{i}\right)$.
Based on Proposition \ref{prop:Heterogeneous SIS stability condition},
we can state the budget-constrained problem as the following optimization
program:

\emph{
\begin{align}
\underset{{\scriptscriptstyle }}{\mbox{max. }} & _{\varepsilon,D,B,W_{\mathcal{G}}}\:\varepsilon\label{eq:Budget-Constrained Spectral Problem}\\
\mbox{s.t. } & \Re\left[\lambda_{1}\left(BW_{\mathcal{G}}-D\right)\right]\leq-\varepsilon,\label{eq:Spectral constraint in budget problem}\\
 & \sum_{i,j}h_{ij}\left(w_{ij}\right)+\sum_{i}\left[f_{i}\left(\beta_{i}\right)+g_{i}\left(\delta_{i}\right)\right]\leq C,\label{eq:Budget constraint in budget problem}\\
 & \underline{\beta}_{i}\leq\beta_{i}\leq\overline{\beta}_{i},\mbox{ }v_{i}\in\mathcal{V},\label{eq:Square constraint for beta in budget problem}\\
 & \underline{\delta}_{i}\leq\delta_{i}\leq\overline{\delta}_{i},\mbox{ }v_{i}\in\mathcal{V},\label{eq:Square constraint for delta in budget problem}\\
 & \underline{w}_{ij}\leq w_{ij}\leq\overline{w}_{ij},\mbox{ }\left(v_{j},v_{i}\right)\in\mathcal{E},\label{eq:Square constraint for traffic}
\end{align}
}where constraint (2.4) forces $\varepsilon$ to be the exponential
decay rate, (\ref{eq:Budget constraint in budget problem}) is the
budget constraint, and (2.6)-(2.8) are the feasible ranges for the
control decision variables. In the following section, we propose an
approach to solve these problems in polynomial time for weighted and
directed contact networks, for a wide class of cost functions $f_{i}$,
$g_{i}$, and $h_{ij}$.

\section{\label{sec:Convex Framework}Geometric Programming for Traffic Control}

The budget-constrained for weighted, directed networks can be solved
using \emph{geometric programming (GP)} \cite{BV04}. Before we present
the details of our solution, we provide some background on GP (a thorough
treatment of GP can be found in \cite{BKVH07}).

\subsection{Geometric Programming Background.}

Let $x_{1},\ldots,x_{n}>0$ denote $n$ decision variables and define
the vector $\mathbf{x}:=\left(x_{1},\ldots,x_{n}\right)\in\mathbb{R}_{++}^{n}$.
In the context of GP, a \emph{monomial $h(\mathbf{x})$} is defined
as a real-valued function of the form $h(\mathbf{x}):=dx_{1}^{a_{1}}x_{2}^{a_{2}}\ldots x_{n}^{a_{n}}$
with $d>0$ and $a_{i}\in\mathbb{R}$. A \emph{posynomial} function
$q(\mathbf{x})$ is defined as a sum of monomials, i.e., $q(\mathbf{x})\triangleq\sum_{k=1}^{K}c_{k}x_{1}^{a_{1k}}x_{2}^{a_{2k}}\ldots x_{n}^{a_{nk}}$,
where $c_{k}>0$. Posynomials are closed under addition, multiplication,
and nonnegative scaling. A posynomial can be divided by a monomial,
with the result a posynomial.

In our formulation, it is useful to define the following class of
functions:

\begin{definition}A function $F:\mathbb{R}^{n}\to\mathbb{R}$ is
\emph{convex in log-scale} if the function 
\begin{equation}
F\left(\mathbf{y}\right)\triangleq\log f\left(\exp\mathbf{y}\right),\label{eq:Convex in Log Scale}
\end{equation}
is convex in \textbf{$\mathbf{y}$} (where $\exp\mathbf{y}$ indicates
component-wise exponentiation).

\end{definition}

Note that posynomials (hence, also monomials) are convex in log-scale
\cite{BV04}.

A geometric program (GP) is an optimization problem of the form (see
\cite{BKVH07} for a comprehensive treatment):
\begin{align}
\mbox{minimize } & f(\mathbf{x})\label{eq:General GP}\\
\mbox{subject to } & q_{i}(\mathbf{x})\leq1,\: i=1,...,m,\nonumber \\
 & h_{i}(\mathbf{x})=1,\: i=1,...,p,\nonumber 
\end{align}
where $q_{i}$ are posynomial functions, $h_{i}$ are monomials, and
$f$ is a convex function in log-scale%
\footnote{Geometric programs in standard form are usually formulated assuming
$f\left(\mathbf{x}\right)$ is a posynomial. In our formulation, we
assume that $f\left(\mathbf{x}\right)$ is in the broader class of
convex functions in logarithmic scale.%
}. A GP is a quasiconvex optimization problem \cite{BV04} that can
be converted into a convex problem. This conversion is based on the
logarithmic change of variables $y_{i}=\log x_{i}$, and a logarithmic
transformation of the objective and constraint functions (see \cite{BKVH07}
for details on this transformation). After this transformation, the
GP in (\ref{eq:General GP}) takes the form
\begin{align}
\mbox{minimize } & F\left(\mathbf{y}\right)\label{eq:Transformed GP}\\
\mbox{subject to } & Q_{i}\left(\mathbf{y}\right)\leq1,\: i=1,...,m,\nonumber \\
 & \mathbf{b}_{i}^{T}\mathbf{y}+\log d_{i}=0,\: i=1,...,p,\nonumber 
\end{align}
where $Q_{i}\left(\mathbf{y}\right)\triangleq\log q_{i}(e^{\mathbf{y}})$,
$F\left(\mathbf{y}\right)\triangleq\log f\left(\exp\mathbf{y}\right)$,
and $\mathbf{b}_{i}=\left(b_{1,i}\ldots b_{n,i}\right)^{T}$. Notice
that, since $f\left(\mathbf{x}\right)$ is convex in log-scale, $F\left(\mathbf{y}\right)$
is a convex function. Also, since $q_{i}$ is a posynomial (therefore,
convex in log-scale), $Q_{i}$ is also a convex function. In conclusion,
(\ref{eq:Transformed GP}) is a convex optimization problem in standard
form and can be efficiently solved in polynomial time \cite{BV04}.

As we shall show in Subsections \ref{sub:GP for SC Digraphs}, we
can solve Problem \ref{Problem: Budget Constrained Allocation} using
GP if the cost functions $f_{i},g_{i},h_{ij}$ are convex in log-scale.
In practice, we model the cost functions as posynomials (see \cite{BKVH07},
Section 8, for a treatment about the modeling abilities of monomials
and posynomials), which are always convex in log-scale.

\subsection{Traffic Control in Directed Networks.\label{sub:GP for SC Digraphs}}

We transform Problem \ref{Problem: Budget Constrained Allocation}
into a GP using elements from the theory of nonnegative matrices.
In our derivations, we use Perron-Frobenius lemma \cite{meyer2000matrix}:

\begin{lemma}\label{lem:Perron-Frobenius}(Perron-Frobenius) Let
$M$ be a nonnegative, irreducible matrix. Then, the following statements
about its spectral radius, $\rho\left(M\right)$, hold:

\emph{(}a\emph{)} $\rho\left(M\right)>0$ is a simple eigenvalue of
$M$,

\emph{(}b\emph{)} $M\mathbf{u}=\rho\left(M\right)\mathbf{u}$, for
some $\mathbf{u}\in\mathbb{R}_{++}^{n}$, and

\emph{(}c\emph{)} $\rho\left(M\right)=\inf\left\{ \lambda\in\mathbb{R}:M\mathbf{u}\leq\lambda\mathbf{u}\mbox{ for }\mathbf{u}\in\mathbb{R}_{++}^{n}\right\} $.

\end{lemma}

Since a matrix $M$ is \emph{irreducible} if and only if its associated
digraph $\mathcal{G}_{M}$ is strongly connected, the above lemma
also holds for the spectral radius of the adjacency matrix of any
(positively) weighted, strongly connected digraph.

From Lemma \ref{lem:Perron-Frobenius}, we infer the following results:

\begin{corollary}\label{cor:Eig equals Rad}Let $M$ be a nonnegative,
irreducible matrix. Then, its eigenvalue with the largest real part,
$\lambda_{1}\left(M\right)$, is real, simple, and equal to the spectral
radius $\rho\left(M\right)>0$.

\end{corollary}

\begin{lemma}

\label{lem:Monotonicity of lambda1}Consider the adjacency matrix
$A_{\mathcal{G}}$ of a (positively) weighted, directed, strongly
connected graph $\mathcal{G}$, and two sets of positive numbers $\left\{ \beta_{i}\right\} _{i=1}^{n}$
and $\left\{ \delta_{i}\right\} _{i=1}^{n}$. Then, $\lambda_{1}\left(\mbox{diag}\left(\beta_{i}\right)A-\mbox{diag}\left(\delta_{i}\right)\right)$
is an increasing function w.r.t. $\beta_{k}$ (respectively, monotonically
decreasing w.r.t. $\delta_{k}$) for $k=1,\ldots,n$.

\end{lemma}

\begin{flushleft}
\textbf{Proof}. In the Appendix.
\par\end{flushleft}

\medskip{}

From the above results, we have the following result (\cite{BV04},
Chapter 4):

\begin{proposition}\label{prop:From PF to Posynomials}Consider the
$n\times n$ nonnegative, irreducible matrix $M\left(\mathbf{x}\right)$
with entries being either $0$ or posynomials with domain $\mathbf{x}\in\mathcal{S}\subseteq\mathbb{R}_{++}^{k}$,
where $\mathcal{S}$ is defined as $\mathcal{S}=\bigcap_{i=1}^{m}\left\{ \mathbf{x\in\mathbb{R}}_{++}^{k}:f_{i}\left(\mathbf{x}\right)\leq1\right\} $,
$f_{i}$ being posynomials. Then, we can minimize $\lambda_{1}\left(M\left(\mathbf{x}\right)\right)$
for $\mathbf{x}\in\mathcal{S}$ solving the following GP:
\begin{align}
\underset{\lambda,\left\{ u_{i}\right\} _{i=1}^{n},\mathbf{x}}{\mbox{minimize }} & \lambda\label{eq:GP for spectral objective}\\
\mbox{subject to } & \frac{\sum_{j=1}^{n}M_{ij}\left(\mathbf{x}\right)u_{j}}{\lambda u_{i}}\leq1,\: i=1,\ldots,n,\label{eq:GP for spectral constraint}\\
 & f_{i}\left(\mathbf{x}\right)\leq1,\: i=1,\ldots,m.
\end{align}
\end{proposition}

Based on the above results, we provide the following solution to the
budget-constrained problems, assuming that the cost functions $f_{i}$
and $g_{i}$ are posynomials and the contact graph $\mathcal{G}$
is strongly connected:

\begin{theorem}\label{thm:GP for budget constrained}Consider the
following elements:
\begin{enumerate}
\item A strongly connected graph \emph{$\mathcal{G}=\left(\mathcal{V},\mathcal{E},\mathcal{W}\right)$,}
\item Posynomial cost functions\emph{$\left\{ f_{i}\left(\beta_{i}\right),g_{i}\left(\delta_{i}\right)\right\} _{v_{i}\in\mathcal{V}}$
and $\left\{ h_{ij}\left(w_{ij}\right)\right\} _{(v_{j},v_{i})\in\mathcal{E}}$,}
\item Bounds on the infection, recovery, and traffic rates $0<\underline{\beta}_{i}\leq\beta_{i}\leq\overline{\beta}_{i}$,
$0<\underline{\delta}_{i}\leq\delta_{i}\leq\overline{\delta}_{i}$,
$i\in\mathcal{V}$,\emph{ }and $0<\underline{w}_{ij}\leq w_{ij}\leq\overline{w}_{ij}$,
\emph{$(v_{j},v_{i})\in\mathcal{E}$,} and
\item A maximum budget $C$ to invest in protection resources.
\end{enumerate}
Then, the optimal investment on vaccines and antidotes for node $v_{i}$
to solve Problem \ref{Problem: Budget Constrained Allocation} are
$f_{i}\left(\beta_{i}^{*}\right)$ and $g_{i}\left(\overline{\Delta}+1-\widehat{\delta}_{i}^{*}\right)$,
where\emph{ $\overline{\Delta}\triangleq\max\left\{ \overline{\delta}_{i}\right\} _{i=1}^{n}$
and }$\beta_{i}^{*}$,$\widehat{\delta}_{i}^{*}$ are the optimal
solution for $\beta_{i}$ and $\widehat{\delta}_{i}$ in the following
GP\emph{:
\begin{align}
\underset{{\scriptstyle }}{\mbox{min.}} & _{\lambda,u_{i},\beta_{i},\widehat{\delta}_{i},t_{i},w_{ij}}\:\lambda\label{eq:Budget-Constrained Spectral Problem-1}\\
\mbox{s.t. } & \frac{\beta_{i}\sum_{j=1}^{n}w_{ij}u_{j}+\widehat{\delta}_{i}u_{i}}{\lambda u_{i}}\leq1,\mbox{ }v_{i}\in\mathcal{V},\label{eq:Eigenvalu condition in spectral constraint}\\
 & \sum_{i,j}h_{ij}\left(w_{ij}\right)+\sum_{i}\left[f_{i}\left(\beta_{i}\right)+g_{i}\left(t_{i}\right)\right]\leq C,\label{eq:budget constraint in budget constrained}\\
 & t_{i}+\widehat{\delta}_{i}\leq\overline{\Delta}+1,\label{eq:t trick in budget constrained}\\
 & \overline{\Delta}+1-\overline{\delta}_{i}\leq\widehat{\delta}_{i}\leq\overline{\Delta}+1-\underline{\delta}_{i},\label{eq:Beta contraint in budget constrained}\\
 & \underline{\beta}_{i}\leq\beta_{i}\leq\overline{\beta}_{i},\mbox{ }v_{i}\in\mathcal{V},\label{eq:Delta constraint in budget constraint}\\
 & \underline{w}_{ij}\leq w_{ij}\leq\overline{w}_{ij},\mbox{ }\left(v_{j},v_{i}\right)\in\mathcal{E},
\end{align}
}

\end{theorem}

\textbf{Proof}. First, based on Proposition \ref{prop:From PF to Posynomials},
we have that maximizing $\varepsilon$ in (\ref{eq:Budget-Constrained Spectral Problem})
subject to (\ref{eq:Spectral constraint in budget problem})-(\ref{eq:Square constraint for beta in budget problem})
is equivalent to minimizing $\lambda_{1}\left(BA_{\mathcal{G}}-D\right)$
subject to (\ref{eq:Budget constraint in budget problem}) and (\ref{eq:Square constraint for beta in budget problem}),
where $B\triangleq\mbox{diag}\left(\beta_{i}\right)$ and $D\triangleq\mbox{diag}\left(\delta_{i}\right)$.
Let us define $\widehat{D}\triangleq\mbox{diag}\left(\widehat{\delta}_{i}\right)$,
where $\widehat{\delta}_{i}\triangleq\overline{\Delta}+1-\delta_{i}$
and $\overline{\Delta}\triangleq\max\left\{ \overline{\delta}_{i}\right\} _{i=1}^{n}$.
Then, $\lambda_{1}\left(BA_{\mathcal{G}}+\widehat{D}\right)=\lambda_{1}\left(BA_{\mathcal{G}}-D\right)+\overline{\Delta}+1$.
Hence, minimizing $\lambda_{1}\left(BA_{\mathcal{G}}-D\right)$ is
equivalent to minimizing $\lambda_{1}\left(BA_{\mathcal{G}}+\widehat{D}\right)$.
The matrix $BA_{\mathcal{G}}+\widehat{D}$ is nonnegative and irreducible
if $A_{\mathcal{G}}$ is the adjacency matrix of a strongly connected
digraph. Therefore, applying Proposition \ref{prop:From PF to Posynomials},
we can minimize $\lambda_{1}\left(BA_{\mathcal{G}}+\widehat{D}\right)$
by minimizing the cost function in (\ref{eq:Budget-Constrained Spectral Problem-1})
under the constraints (\ref{eq:Eigenvalu condition in spectral constraint})-(\ref{eq:Delta constraint in budget constraint}).
Constraints (\ref{eq:Beta contraint in budget constrained}) and (\ref{eq:Delta constraint in budget constraint})
represent bounds on the achievable infection and curing rates. Notice
that we also have a constraint associated to the budget available,
i.e., $\sum_{k=1}^{n}f_{k}\left(\beta_{k}\right)+g_{k}\left(\overline{\Delta}+1-\widehat{\delta}_{i}\right)\leq C$.
But, even though $g_{k}$$\left(\delta_{k}\right)$ is a polynomial
function on $\delta_{k}$, $g_{k}\left(\overline{\Delta}+1-\widehat{\delta}_{k}\right)$
is not a posynomial on $\widehat{\delta}_{i}$. To overcome this issue,
we can replace the argument of $g_{k}$ by a new variable $t_{k}$,
along with the constraint $t_{k}\leq\overline{\Delta}+1-\widehat{\delta}_{k}$,
which can be expressed as the posynomial inequality, $\left(t_{k}+\widehat{\delta}_{k}\right)/\left(\overline{\Delta}+1\right)\leq1$.
As we proved in Lemma \ref{lem:Monotonicity of lambda1}, the largest
eigenvalue $\lambda_{1}\left(BA-D\right)$ is a decreasing function
of $\delta_{k}$ and the antidote cost function $g_{k}$ is monotonically
increasing w.r.t. $\delta_{k}$. Thus, adding the inequality $t_{k}\leq\overline{\Delta}+1-\widehat{\delta}_{k}$
does not change the result of the optimization problem, since at optimality
$t_{k}$ will saturate to its largest possible value $t_{k}=\overline{\Delta}+1-\widehat{\delta}_{k}$.

\begin{flushright}
$\blacksquare$
\par\end{flushright}

\section{\label{sub:Numerical-Results}Numerical Results}

We analyze data from a real-world air transportation network and find
the optimal traffic control strategy to prevent the viral spreading
of an epidemic outbreak that propagates through the air transportation
network \cite{schneider2011suppressing}. For clarity in our exposition,
we limit our control actions to regulate traffic in the edges of the
air traffic network, keeping the infection and recovery rates fixed
(although we could use the GP in Theorem \ref{thm:GP for budget constrained}
to control, simultaneously, traffic, prevention and correction resources).
The air transportation network under analysis spans the major airports
in the world, in particular, those having an incoming traffic greater
than 10 million passengers per year (MPPY). There are $56$ such airports
world-wide and they are connected via $1,843$ direct flights, which
we represent as directed edges in a graph weighted graph. The weight
of each directed edge represents the number of passengers taking that
flight throughout the year (in MPPY units).

In our simulations, we consider the following values for the infection
and recovery rate: $\delta_{i}=0.1$ and $\beta_{i}=0.033$, for $i=1,\ldots,56$.
In the absence of traffic-control resources, the matrix $BA_{\mathcal{G}}-D$
in (\ref{eq:Spectral Control}) has its largest eigenvalue at $\lambda_{1}\left(\beta_{i}A_{\mathcal{G}}-\delta_{i}I\right)=0.21>0$;
thus, the disease-free equilibrium is unstable and a random initial
infection can propagate through the air transportation network. We
now find the optimal allocation of traffic control resources to stabilize
the disease-free equilibrium to protect the population against an
epidemic outbreak propagating through the air traffic infrastructure.

\begin{figure}
\centering\includegraphics[width=0.95\columnwidth]{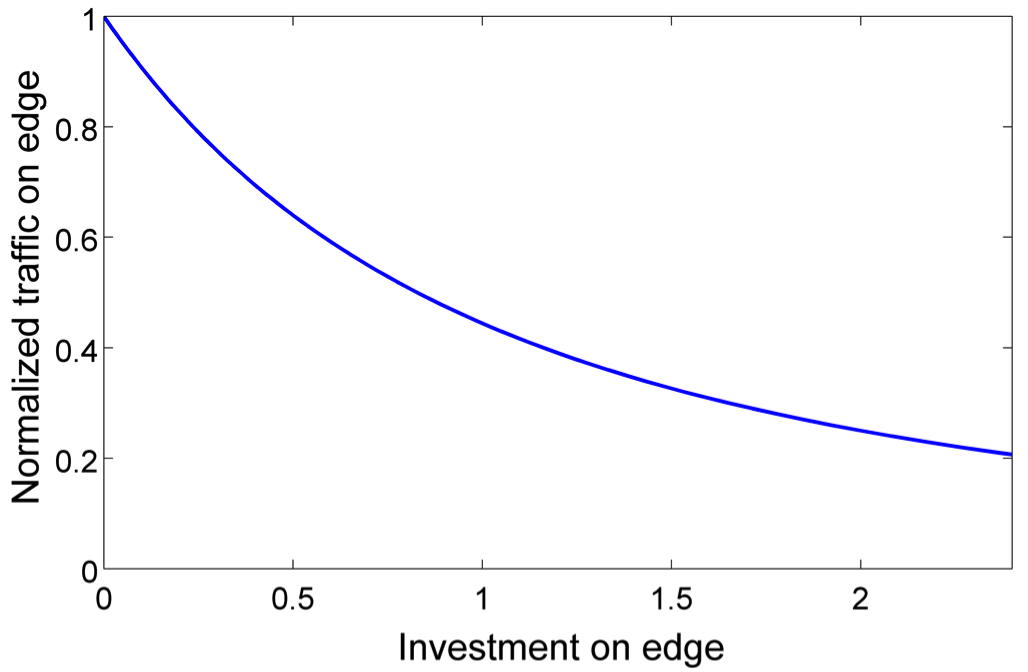}\caption{Traffic rate achieved at edge $\left(v_{j},v_{i}\right)$ after an
investment on traffic control (in abscissas) is made on that edge.}
\label{fig. cost functions}
\end{figure}

In our simulations, we consider the case in which we can control the
traffic flowing in a particular directed edge by investing on protection
resources on that edge. For example, the authority responsible for
traffic management can decide to reduce the number of passengers flying
in a particular flight. This measure has the cost of compensating
those passengers forced to miss their flight. In our simulations,
we consider the following traffic cost functions:
\begin{equation}
h_{ij}\left(w_{ij}\right)=p\left(w_{ij}^{-1/p}-\overline{w}_{ij}^{-1/p}\right).\label{eq:Quasiconvex Limit-1}
\end{equation}
In Fig. \ref{fig. cost functions}, we plot this cost function for
$\overline{w}_{ij}=1$ and $p=2$, where the abscissa is the amount
invested in traffic control on a particular edge and the ordinates
are the traffic rate achieved by the investment. Notice that in the
absence of investment, the achieved traffic is $\overline{w}_{ij}$.
As we increase the amount invested on traffic control on a particular
edge $\left(v_{j},v_{i}\right)$, the traffic rate $w_{ij}$ of that
edge is reduced. Notice that the cost function chosen in our experiment
presents diminishing marginal benefit on investment. Moreover, we
also impose a lower bound on the amount of traffic allowed in each
flight. In particular, the competent authority can force up to $80\%$
of the passengers of a flight to stay on the ground. Therefore, the
lowest possible amount of traffic on an edge corresponds to $\underline{w}_{ij}:=0.2\overline{w}_{ij}$.

\textbf{}%

\begin{figure*}
\centering\includegraphics[width=1\textwidth]{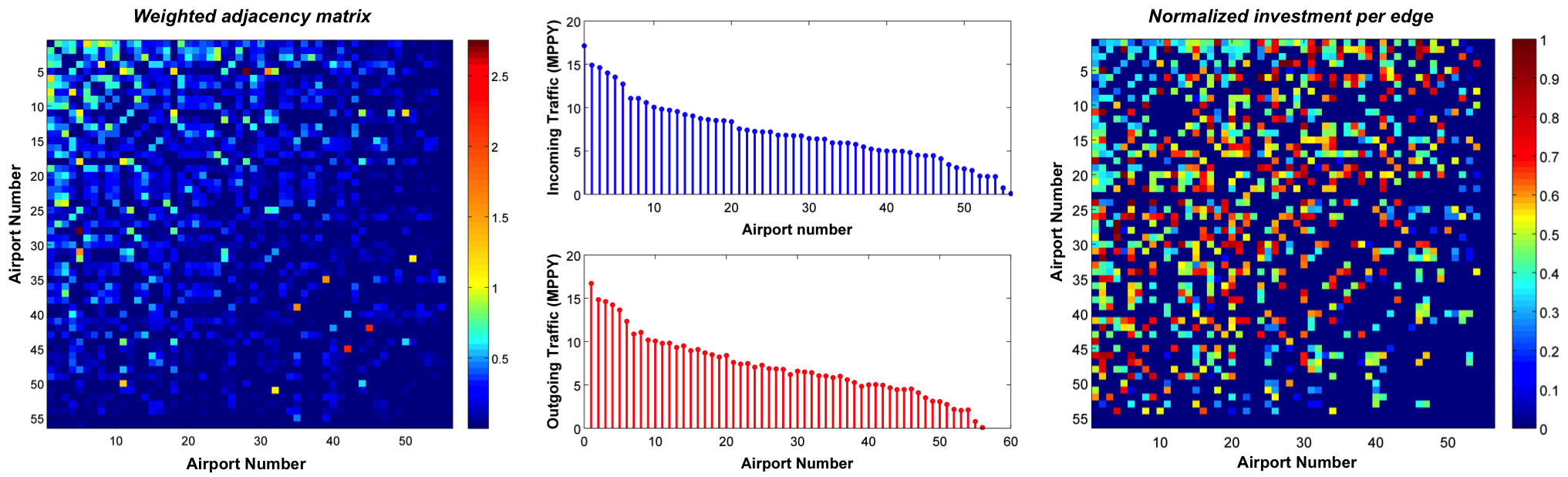}
\caption{In the left subplot, we represent the colormap of the adjacency matrix
of the air transportation network under study. In the middle plots
of Fig. \ref{fig_rate-constrained}, we represent the incoming traffic
(above) and out-going traffic (below) for each one of the airports
under consideration. In the right subplot, we include a colormap corresponding
to the optimal allocation of traffic-control resources over the set
of directed edges.}

\label{fig_rate-constrained} 
\end{figure*}

Using the air transportation network, the parameters, and the cost
functions specified above, we solve both the budget-constrained allocation
problem using the geometric program in Theorems \ref{thm:GP for budget constrained}.
The solution of the budget-constrained allocation problem is summarized
in Fig. \ref{fig_rate-constrained}. In the left subplot, we represent
the colormap of the adjacency matrix of the air transportation network
under study. The color of each pixel in this plot corresponds to the
value of $\overline{w}_{ij}$, the maximum achievable traffic in each
edge, measured in MPPY. To each one of the 56 airports under study,
we have associated a number which corresponds to its ranking with
respect to incoming traffic in the airport. In the middle plots of
Fig. \ref{fig_rate-constrained}, we represent the incoming traffic
(above) and out-going traffic (below) for each one of the airports
under consideration.

Using Theorem \ref{thm:GP for budget constrained}, we solve the budget-constrained
allocation problem with a total budget of 300 units. With this budget,
we achieve an optimal exponential decay rate of $\varepsilon^{*}=0.021$.
The corresponding allocation of traffic-control resources over the
set of directed edges is summarized in the right subplot of Fig. \ref{fig_rate-constrained}.
The color of the $\left(i,j\right)$ pixel in the colormap represents
the amount of control resources invested on edge $\left(v_{j},v_{i}\right)$.
A dark blue pixel corresponds to the absence of investment on controlling
that edge; thus, the traffic through that edge is not modified and
the flow of passengers is equal to $\overline{w}_{ij}$ (the maximum
possible flow). A dark red pixel corresponds to a normalized value
of investment equal to one. This value corresponds to a saturation
of traffic control resources in that edge. In other words, a dark
red $\left(i,j\right)$ pixel indicates that the flow of passengers
through edge $\left(v_{j},v_{i}\right)$ has been reduced to the minimum
possible value, which we have chosen to be $\underline{w}_{ij}=0.2\overline{w}_{ij}$.
Notice how, the optimal traffic pattern represented in the right subplot
indicates a nontrivial distribution of resources throughout the edges
of the transportation network.

\begin{figure*}
\centering\includegraphics[width=1\textwidth]{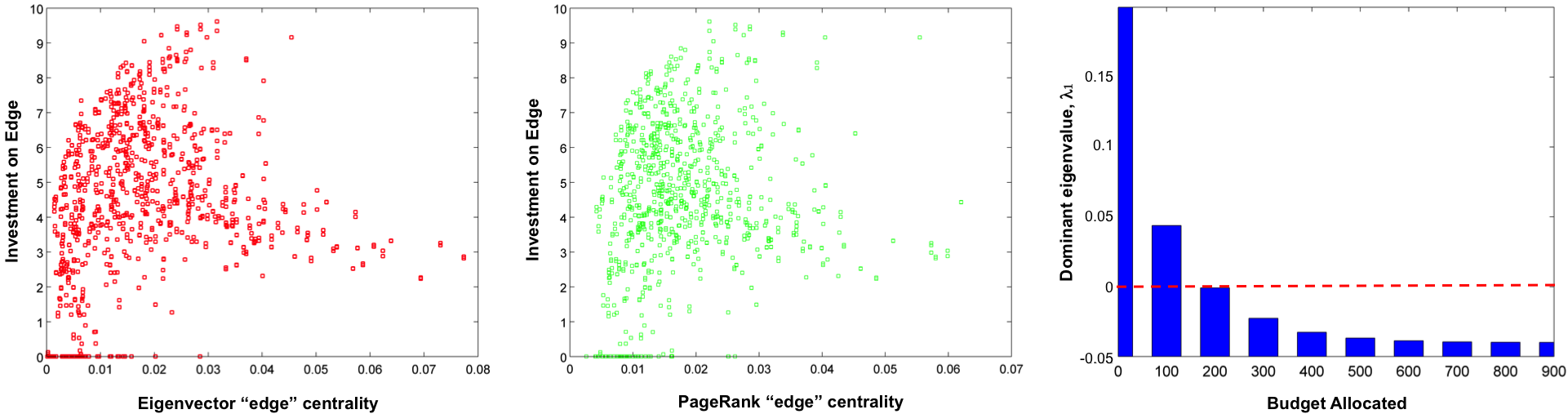}
\caption{Results from the budget-constrained allocation problem. From left
to right, we have (\emph{a}) a scatter plot with the investment on
traffic control versus the eigenvector centrality of each edge, (\emph{b})
a scatter plot with the investment on traffic control versus the PageRank
centrality of each edge, and (\emph{c}) bar plot describing the variation
in the achieved rate of containment as a function of the total investment
budget.}

\label{fig_budget-constrained} 
\end{figure*}

The resulting pattern of investment on traffic-control resources is
not trivially related with any of the centrality measures popularly
considered in the literature. In Fig. \ref{fig_budget-constrained},
we plot the relationship between the amount invested on an edge and
a measure of the edge centrality. Although there are some measures
of edge centrality in the literature, the concept of node centrality
is better developed. We consider in our illustrations two measures
of edge centrality based on the centralities of the nodes connected
by the edge. In particular, we consider both the eigenvector and the
PageRank centralities of the nodes in the network \cite{New10}. Denoting
by $v_{i}$ and $r_{i}$ the eigenvector and the PageRank centralities
of node $v_{i}\in\mathcal{V}$, we define the corresponding eigenvector
and PageRank centrality of an edge $\left(v_{j},v_{i}\right)$ as
$v_{ij}=v_{i}v_{j}$ and $r_{ij}=r_{i}r_{j}$. In the left and center
subplots in Fig. \ref{fig_budget-constrained}, we include two scatter
plots where each point represents an edge in the transportation network.
The abscissas in those plots are the eigenvector and PageRank centralities
of each edge, $v_{ij}$ and $r_{ij}$ respectively, and the ordinates
are the amount of traffic-control resources invested on that edge.
We observe that there is no trivial law relating the optimal investment
on an edge with these edge centrality measurements. For example, we
observe in Fig. \ref{fig_budget-constrained} how some flights connecting
airports of low centrality receive higher investment on protection
than other flights connecting airports with higher centrality.

Finally, we also included in the right subplot of Fig. \ref{fig_budget-constrained}
a bar plot describing the variation in the achieved rate of containment
as a function of the total investment budget. Observe how, for initial
values of investment, we achieve a drastic decrease in the containment
rate $\lambda_{1}$. This decrease flattens out as we increase the
total budget allocated, since the traffic cost function, $h_{ij}$,
used in our simulations presents a diminishing marginal benefit on
investment. We also observe that the bar diagram becomes completely
flat for a level of investment over 900 monetary units. This level
corresponds to a saturated level of containments in which all flights
have been controlled to have a flow equal to $w_{ij}=\underline{w}_{ij}=0.2\overline{w}_{ij}$.

{}

\section{\label{sec:Conclusions}Conclusions}

The problem of allocating protection resources to contain spreading
processes has been studied for the case of weighted, directed networks.
Relevant applications include the propagation of viruses in computer
networks, cascading failures in complex technological networks, and
the spreading of epidemics in human populations. Three types of resources
has been considered: (\emph{i}) Traffic control resources which constrain
the flow over edges in the contact graph, (\emph{ii}) \emph{preventive}
resources which `immunize' nodes against the spreading (e.g. vaccines),
and (\emph{iii}) \emph{corrective} resources which neutralize the
infection after it has reached a node (e.g. antidotes). We assume
that all resource types have an associated cost (which may vary between
nodes or edges). Using the \emph{budget-constrained allocation problem},
we have found the optimal allocation of resources that contains the
spreading process with a fixed budget. Our solution is built on a
convex optimization framework, specifically Geometric Programming
(GP), which allows us to solve this problem \emph{exactly}--without
relaxations or heuristics--for \emph{weighted and directed} networks
of \emph{nonidentical} agents in polynomial time. A key feature of
the GP approach is that resource allocations of all three types are
optimized \emph{simultaneously}, even in the case where the resource
cost functions are heterogeneous throughout the network.

We have demonstrated our optimal protection strategy for the case
of a hypothetical world-wide pandemic, spread via the air transportation
network. The study has been limited to the airports with the highest
passenger traffic worldwide. Given this network, we have computed
the optimal resource allocation to protect against such an epidemic.
Our simulations indicate nontrivial traffic restriction and resource
allocation patterns which cannot, in general, be captured using simple
heuristics.

\section*{Appendix.}

\textbf{Proof of Lemma \ref{lem:Monotonicity of lambda1}.} We define
the auxiliary matrix $M\triangleq\mbox{diag}\left(\beta_{i}\right)A-\mbox{diag}\left(\delta_{i}\right)+\Delta I$,
where $\Delta\triangleq\max\left\{ \delta_{i}\right\} $. Thus, $\lambda_{1}\left(M\right)=\lambda_{1}\left(\mbox{diag}\left(\beta_{i}\right)A-\mbox{diag}\left(\delta_{i}\right)\right)+\Delta$.
Notice that both $M$ and $M^{T}$ are nonnegative and irreducible
if $\mathcal{G}$ is strongly connected. Hence, from Lemma \ref{lem:Perron-Frobenius},
there are two positive vectors $\mathbf{v}$ and $\mathbf{w}$ such
that
\begin{align*}
M\mathbf{v} & =\rho\mathbf{v},\\
\mathbf{w}^{T}M & =\rho\mathbf{w}^{T},
\end{align*}
where $\rho=\rho\left(M\right)=\lambda_{1}\left(M\right)$, and $\mathbf{v}$,
$\mathbf{w}$ are the right and left dominant eigenvectors of $M$.
From eigenvalue perturbation theory, we have that the increment in
the spectral radius of $M$ induced by a matrix increment $\Delta M$
is \cite{trefethen1997numerical}
\begin{equation}
\rho\left(M+\Delta M\right)-\rho\left(M\right)=\mathbf{w}^{T}\Delta M\mathbf{v}+o\left(\left\Vert \Delta M\right\Vert \right).\label{eq:Eigenvalue Perturbation}
\end{equation}

To study the effect of a positive increment in $\beta_{k}$ in the
spectral radius, we define $\Delta B=\Delta\beta_{k}\mathbf{e}_{k}\mathbf{e}_{k}^{T}$,
for $\Delta\beta_{k}>0$, and apply \ref{eq:Eigenvalue Perturbation}
with $\Delta M=\Delta B\, A$. Hence,
\begin{align*}
\rho\left(M+\Delta M\right)-\rho\left(M\right) & =\Delta\beta_{k}\mathbf{w}^{T}\mathbf{e}_{k}\mathbf{e}_{k}^{T}A\mathbf{v}+o\left(\left\Vert \Delta\beta_{k}\right\Vert \right)\\
 & =\Delta\beta_{k}w_{k}\mathbf{a}_{k}^{T}\mathbf{v}+o\left(\left\Vert \Delta\beta_{k}\right\Vert \right)>0,
\end{align*}
where $\mathbf{a}_{k}^{T}=\mathbf{e}_{k}^{T}A$ and the last inequality
if a consequence of $\Delta\beta_{k}$, $w_{k}$, and $\mathbf{a}_{k}^{T}\mathbf{v}$
being all positive. Hence, a positive increment in $\beta_{k}$ induces
a positive increment in the spectral radius.

\begin{flushright}
Similarly, to study the effect of a positive increment in $\delta_{k}$
in the spectral radius, we define $\Delta D=\Delta\delta_{k}\mathbf{e}_{k}\mathbf{e}_{k}^{T}$,
for $\Delta\delta_{k}>0$. Applying \ref{eq:Eigenvalue Perturbation}
with $\Delta M=-\Delta D$, we obtain
\begin{align*}
\rho\left(M+\Delta D\right)-\rho\left(M\right) & =-\Delta\delta_{k}\mathbf{w}^{T}\mathbf{e}_{k}\mathbf{e}_{k}^{T}\mathbf{v}+o\left(\left\Vert \Delta\delta_{k}\right\Vert \right)\\
 & =-\Delta\delta_{k}w_{k}v_{k}+o\left(\left\Vert \Delta\delta_{k}\right\Vert \right)<0.
\end{align*}
$\blacksquare$
\par\end{flushright}

\bibliographystyle{ieeetr}
\bibliography{ViralSpread}

\begin{thebibliography}{10}

\bibitem{Bai75}
N.~Bailey, {\em The mathematical theory of infectious diseases and its
  applications}.
\newblock Charles Griffin \& Company Ltd., 1975.

\bibitem{GGT03}
M.~Garetto, W.~Gong, and D.~Towsley, ``Modeling malware spreading dynamics,''
  in {\em IEEE INFOCOM 2003}, vol.~3, pp.~1869--1879, 2003.

\bibitem{roy2012security}
S.~Roy, M.~Xue, and S.~K. Das, ``Security and discoverability of spread
  dynamics in cyber-physical networks,'' {\em IEEE Transactions on Parallel and
  Distributed Systems}, vol.~23, no.~9, pp.~1694--1707, 2012.

\bibitem{watts2005multiscale}
D.~J. Watts, R.~Muhamad, D.~C. Medina, and P.~S. Dodds, ``Multiscale, resurgent
  epidemics in a hierarchical metapopulation model,'' {\em Proceedings of the
  National Academy of Sciences}, vol.~102, no.~32, pp.~11157--11162, 2005.

\bibitem{AM91}
R.~Anderson, R.~May, and B.~Anderson, {\em Infectious Diseases of Humans:
  Dynamics and Control}, vol.~28.
\newblock Wiley, 1992.

\bibitem{WCWF03}
Y.~Wang, D.~Chakrabarti, C.~Wang, and C.~Faloutsos, ``Epidemic spreading in
  real networks: An eigenvalue viewpoint,'' in {\em Proc. 22nd Int. Symp.
  Reliable Distributed Systems}, pp.~25--34, 2003.

\bibitem{GMT05}
A.~Ganesh, L.~Massoulie, and D.~Towsley, ``The effect of network topology on
  the spread of epidemics,'' in {\em IEEE INFOCOM 2005}, vol.~2,
  pp.~1455--1466, 2005.

\bibitem{CWWLF08}
D.~Chakrabarti, Y.~Wang, C.~Wang, J.~Leskovec, and C.~Faloutsos, ``Epidemic
  thresholds in real networks,'' {\em ACM Transactions on Information and
  System Security}, vol.~10, no.~4, pp.~1--26, 2008.

\bibitem{MOK09}
P.~Van~Mieghem, J.~Omic, and R.~Kooij, ``Virus spread in networks,'' {\em
  IEEE/ACM Transactions on Networking}, vol.~17, no.~1, pp.~1--14, 2009.

\bibitem{cohen2003efficient}
R.~Cohen, S.~Havlin, and D.~Ben-Avraham, ``Efficient immunization strategies
  for computer networks and populations,'' {\em Physical Review Letters},
  vol.~91, no.~24, p.~247901, 2003.

\bibitem{BCGS10}
C.~Borgs, J.~Chayes, A.~Ganesh, and A.~Saberi, ``How to distribute antidote to
  control epidemics,'' {\em Random Structures \& Algorithms}, vol.~37, no.~2,
  pp.~204--222, 2010.

\bibitem{chung2009distributing}
F.~Chung, P.~Horn, and A.~Tsiatas, ``Distributing antidote using pagerank
  vectors,'' {\em Internet Mathematics}, vol.~6, no.~2, pp.~237--254, 2009.

\bibitem{WRS08}
Y.~Wan, S.~Roy, and A.~Saberi, ``Designing spatially heterogeneous strategies
  for control of virus spread,'' {\em Systems Biology, IET}, vol.~2, no.~4,
  pp.~184--201, 2008.

\bibitem{GOM11}
E.~Gourdin, J.~Omic, and P.~Van~Mieghem, ``Optimization of network protection
  against virus spread,'' in {\em 8th International Workshop on the Design of
  Reliable Communication Networks}, pp.~86--93, 2011.

\bibitem{PZEJP13}
V.~M. Preciado, M.~Zargham, C.~Enyioha, A.~Jadbabaie, and G.~Pappas, ``Optimal
  vaccine allocation to control epidemic outbreaks in arbitrary networks,'' in
  {\em IEEE Conference on Decision and Control}, 2013.

\bibitem{PDS13}
V.~M. Preciado, F.~Darabi~Sahneh, and C.~Scoglio, ``A convex framework for
  optimal investment on disease awareness in social networks,'' in {\em IEEE
  Global Conference on Signal and Information Processing}, 2013.

\bibitem{weiss1971asymptotic}
G.~H. Weiss and M.~Dishon, ``On the asymptotic behavior of the stochastic and
  deterministic models of an epidemic,'' {\em Mathematical Biosciences},
  vol.~11, no.~3, pp.~261--265, 1971.

\bibitem{VO13}
P.~V. Mieghem and J.~Omic, ``In-homogeneous virus spread in networks,'' {\em
  arXiv preprint arXiv:1306.2588}, 2013.

\bibitem{van2006performance}
P.~Van~Mieghem, {\em Performance analysis of communications networks and
  systems}.
\newblock Cambridge University Press, 2006.

\bibitem{BV04}
S.~Boyd and L.~Vandenberghe, {\em Convex optimization}.
\newblock Cambridge university press, 2004.

\bibitem{BKVH07}
S.~Boyd, S.-J. Kim, L.~Vandenberghe, and A.~Hassibi, ``A tutorial on geometric
  programming,'' {\em Optimization and engineering}, vol.~8, no.~1,
  pp.~67--127, 2007.

\bibitem{meyer2000matrix}
C.~D. Meyer, {\em Matrix analysis and applied linear algebra}.
\newblock SIAM, 2000.

\bibitem{schneider2011suppressing}
C.~M. Schneider, T.~Mihaljev, S.~Havlin, and H.~J. Herrmann, ``Suppressing
  epidemics with a limited amount of immunization units,'' {\em Physical Review
  E}, vol.~84, no.~6, p.~061911, 2011.

\bibitem{New10}
M.~Newman, {\em Networks: An introduction}.
\newblock Cambridge University Press, 2010.

\bibitem{trefethen1997numerical}
L.~N. Trefethen and D.~Bau~III, {\em Numerical linear algebra}.
\newblock No.~50, Siam, 1997.

\end{thebibliography}

\end{document}